\documentclass[%
aip,
 amsmath,amssymb,
 reprint,%
]{revtex4-1}

\usepackage{graphicx}
\usepackage{dcolumn}
\usepackage{bm}
\usepackage{color}

\usepackage[utf8]{inputenc}
\usepackage[T1]{fontenc}
\usepackage{mathptmx}
\usepackage{amsmath}
\usepackage{etoolbox}

\makeatletter
\def\@email#1#2{%
 \endgroup
 \patchcmd{\titleblock@produce}
  {\frontmatter@RRAPformat}
  {\frontmatter@RRAPformat{\produce@RRAP{*#1\href{mailto:#2}{#2}}}\frontmatter@RRAPformat}
  {}{}
}%
\makeatother
\begin{document}

\preprint{AIP/123-QED}

\title{Development of ultra-high efficiency soft X-ray angle-resolved photoemission spectroscopy equipped with deep prior-based denoising method}
\author{Kohei Yamagami}
\email{kohei.yamagami@spring8.or.jp}
\affiliation{Japan Synchrotron Radiation Research Institute (JASRI), Sayo, Hyogo 679-5198, Japan.}%
\author{Yuichi Yokoyama}%
\affiliation{Japan Synchrotron Radiation Research Institute (JASRI), Sayo, Hyogo 679-5198, Japan.}%
\author{Yuta Sumiya}%
\affiliation{The University of Electro-Communications, Department of Informatics, Chofu, Tokyo 182-8585, Japan.}%
\author{Hayaru Shouno}%
\affiliation{The University of Electro-Communications, Department of Informatics, Chofu, Tokyo 182-8585, Japan.}%
\author{Tetsuro Nakamura}
\affiliation{Japan Synchrotron Radiation Research Institute (JASRI), Sayo, Hyogo 679-5198, Japan.}%
\author{Masaichiro Mizumaki}%
\affiliation{Faculty of Science, Kumamoto University, Kurokami, Kumamoto 860-8555, Japan.}%

\date{\today}

\begin{abstract}
Soft X-ray angle resolved photoemission spectroscopy (SX-ARPES) is one of the most powerful spectroscopic techniques to visualize the three-dimensional bulk electronic structure in reciprocal lattice space.
Compared with ARPES employing low-energy photon sources, the time burden imposed by a lower photoelectron yield, stemming from the photoionization cross-section, has been a persistent technical challenge.
To address this challenge, we have developed a noise-reduction system by using the deep prior-based approach and integrated it into the micro-focused SX-ARPES ($\mu$SX-ARPES) system at BL25SU in SPring-8.
The implemented system effectively eliminates instrumental artifacts, such as grid and spike structures typical of ARPES data acquired using the voltage Fixed-mode, within approximately 30 s.
We demonstrate, through the $\mu$SX-ARPES measurements on a single crystal of CeRu$_2$Si$_2$, that data with sufficient statistical accuracy can be obtained in approximately 40 s.
In addition, we present the potential of high signal-to-noise ratio ARPES measurement, achieving an energy resolution of 51.6~meV at an excitation energy of 708~eV in $\mu$SX-ARPES measurements on polycrystalline gold.
Our developed system successfully reduces the time burden in SX-ARPES and paves the way for advancements in lower photoelectron yield measurements, such as those requiring higher energy resolution and three-dimensional nonequilibrium measurements.
\end{abstract}

\maketitle
\section{Introduction}
Angle-resolved photoemission spectroscopy (ARPES) has emerged as a powerful experimental technique for directly visualizing the electronic structure of materials as a function of energy and momentum in reciprocal lattice space~\cite{damascelli2003angle,strocov2019k,sobota2021angle,zhang2022angle}.
Understanding the intricate characteristics of band structures, including topological properties, strongly correlated electron effects, and spin splitting arising from symmetry breaking, is crucial for comprehending the fascinating phenomena observed in condensed matter physics.
These phenomena are being progressively explored using vacuum ultraviolet ARPES (VUV-ARPES) with laser and synchrotron light sources~\cite{yang2018visualizing,zhou2018new,li2023recent,boschini2024time}.
In contrast to VUV-ARPES, soft X-ray ARPES (SX-ARPES) excels in revealing three-dimensional fermiology~\cite{sekiyama2000probing,yano2007three,yano2008electronic,strocov2012three,nakatani2018evidence} and probing buried electronic structures, owing to enhancement of bulk sensitivity~\cite{berner2015dimensionality,strocov2019k}.
This sensitivity stems from the fact that the detection depth in SX-ARPES is empirically observed to be 3-5 times larger than that in VUV-ARPES~\cite{seah1979quantitative,tanuma1988calculations,tanuma1991calculations,tanuma1993calculations,tanuma1997calculations,powell2000evaluation}.
Moreover, the capability of zone-selective band separation through the simultaneous detection of multiple Brillouin zones, facilitated by the wider photoelectron detection angle of advanced photoelectron analyzers~\cite{fujiwara2023impact,morita2024zone}, contributes to an accelerated understanding of electronic structures, particularly when employed in conjunction with VUV-ARPES.

\begin{figure*}
\includegraphics[width=17cm]{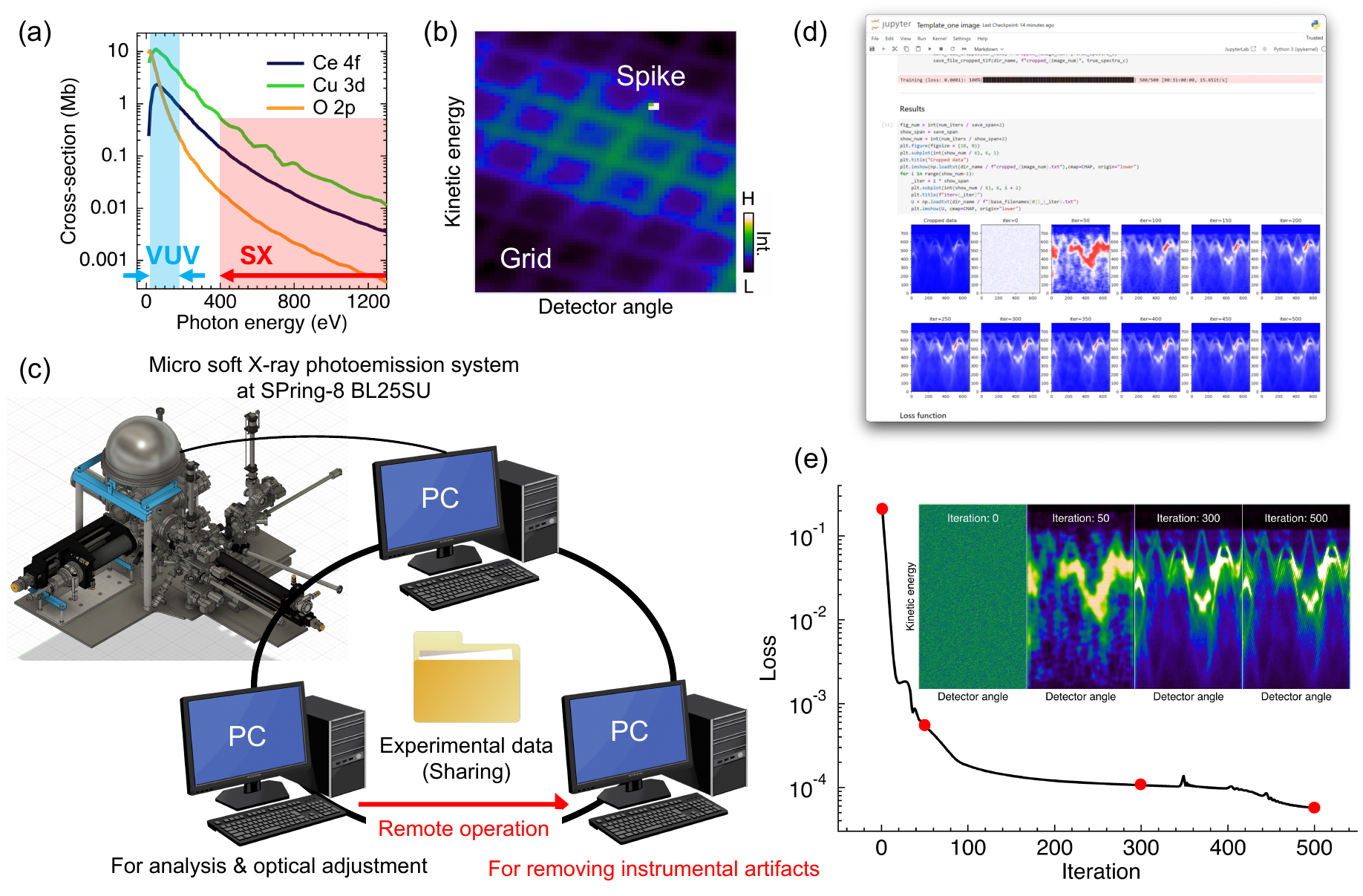}
\caption{\label{fig:epsart} (a) Photoionization cross-sections of Ce 4$f$, Cu 3$d$, and O 2$p$ electrons, digitized from Ref.~\onlinecite{yeh1985atomic}. (b) Instrumental artifacts such as grid and spike structures commonly observed in ARPES image obtained by Fixed mode. (c) Overview of DPDM system implementation. (d) Screenshot of the jupyter Notebook interface after running DPDM. The number of iterations, the degree of completion, and the noise processing time are indicated within the red frame. The ARPES image can be viewed during the denoising process, allowing for real-time monitoring. The optimal denoised data can be selected by referring to the Loss function shown below. (e) Loss function as a function of iterations. The inset ARPES images correspond to the iteration counts indicated by the red circles.}
\label{Fig.1}
\end{figure*}

However, the photoelectron yield in SX-ARPES remains considerably lower than that in VUV-ARPES, despite significant improvements in photon flux and density provided by upgraded low-emittance fourth-generation soft X-ray synchrotron radiation sources~\cite{tavares2018commissioning,fornek2019advanced,raimondi2021commissioning,klysubun2022sps,watanabe2023spring,obara2025commissioning,obara2025commissioning}.
This fundamental constraint arises primarily from the photoionization cross-section~\cite{yeh1985atomic}, which governs the photoemission process.
As depicted in Fig.~\ref{Fig.1}(a), the cross-section for valence electrons of typical elements exhibits a strong dependence on photon energy, with the cross-section in the SX region being more than an order of magnitude smaller than that in the VUV region.
This reduced cross-section requires longer measurement times in SX-ARPES.
Such prolonged acquisition periods risk introducing undesirable effects, such as surface modification or beam drift, which obstruct the accurate determination of the intrinsic electronic structure.

To address the challenge posed by the longer measurement times, the Fixed mode of operation in an electrostatic hemispherical photoelectron analyzer offers a viable solution.
In this mode, photoelectrons are measured within a specific kinetic-energy range, limited to the detection area of a two-dimensional detector.
This detector typically comprises a multichannel plate (MCP), a phosphor screen, and a camera.
Compared to the Swept mode, which involves scanning a wide kinetic energy range, the Fixed mode reduces the acquisition time required to obtain a spectrum with a high statistical signal-to-noise (S/N) ratio.
Moreover, this local measurement mode is particularly well-suited for detailed analysis of the band structure near the Fermi level due to its superior angular resolution.
As reported in a previous study \cite{strocov2014soft}, any slight displacement of the X-ray spot from the electron-optical axis of the analyzer tilts the angle-resolved image on the detector.
In the Swept mode, the summation of these tilted images during energy scanning leads to a smearing of the angular resolution.
The Fixed mode avoids this issue by maintaining a constant kinetic energy, thereby preserving the highest possible angular resolution.
However, a common implementation part in an electrostatic hemispherical photoelectron analyzer introduces a potential issue.
To block stray photoelectrons, a wire mesh is often placed just before the MCP as a type of electron filter~\cite{liu2023removing}.
Stray photoelectrons are electrons traveling in unintended trajectories enter the detector during precisely analysis process, which they contribute as background to the data, thus lowering the spectrum quality.
This mesh leaves traces of instrumental artifacts on the raw data, including a periodic grid structure as well as non-periodic spike structures arising from the aging of the detection unit (Fig.~\ref{Fig.1}(b)).
Conventionally, such artifacts are averaged out through physical scanning methods, such as the Swept mode or dithering (which involves small voltage offsets).
While effective, these physical approaches inherently increase the total acquisition time and the complexity of the measurement system, thereby offsetting the efficiency gains of the Fixed mode.
Consequently, the elimination technique are necessary to remove these artifacts.

Traditional mathematical noise processing methods based on Fourier transforms~\cite{liu2022fourier} are not ideal for addressing aperiodic spike structures.
Moreover, these methods heavily rely on manual adjustments and are sensitive to experimental conditions, potentially compromising the reliability of the processed data.
To overcome these limitations, we have recently developed the Deep Prior-based Denoising Method (DPDM)~\cite{deep_prior}.
This method offers a powerful, training-free approach to image restoration.
The technique employs a four-layer U-shaped convolutional neural network (CNN) with skip connections, leveraging its inherent structural priors alongside an early stopping strategy.
By minimizing the mean squared error (MSE) between the U-Net output and the target ARPES image as a loss function, the network architecture itself acts as a filter that prioritizes natural image features over unnatural noise.
DPDM is also suitable for removing regular grid structures, as the spectral bias~\cite{rahaman2019spectral} ensures that their high-frequency features are not learned until the lower-frequency band structures are completely restored.
This approach effectively separates the intrinsic signal from instrumental artifacts and statistical noise, yielding clearer ARPES images in a fraction of the time required by conventional Swept mode.

This paper reports the development of an ultra-high efficient SX-ARPES system capable of acquiring S/N ARPES images and performing denoising within one minute.
This system integrates a micro-focusing soft X-ray photoelectron spectroscopy system ($\mu$SX-ARPES) operating at BL25SU of SPring-8 with a newly developed DPDM.
Single crystals of CeRu$_2$Si$_2$ and Mn$_3$Si$_2$Te$_6$, known for exhibiting clear and indistinct band dispersions, respectively, were selected as test samples to evaluate the accumulation time dependence and capture buried hand structure from the background facilitated by DPDM.
Our developed system demonstrates that DPDM accelerates the high-throughput capabilities of synchrotron-based SX-ARPES and opens up new possibilities for developing low-yield SX-ARPES measurement techniques.

\begin{figure}
\includegraphics[width=8.5cm]{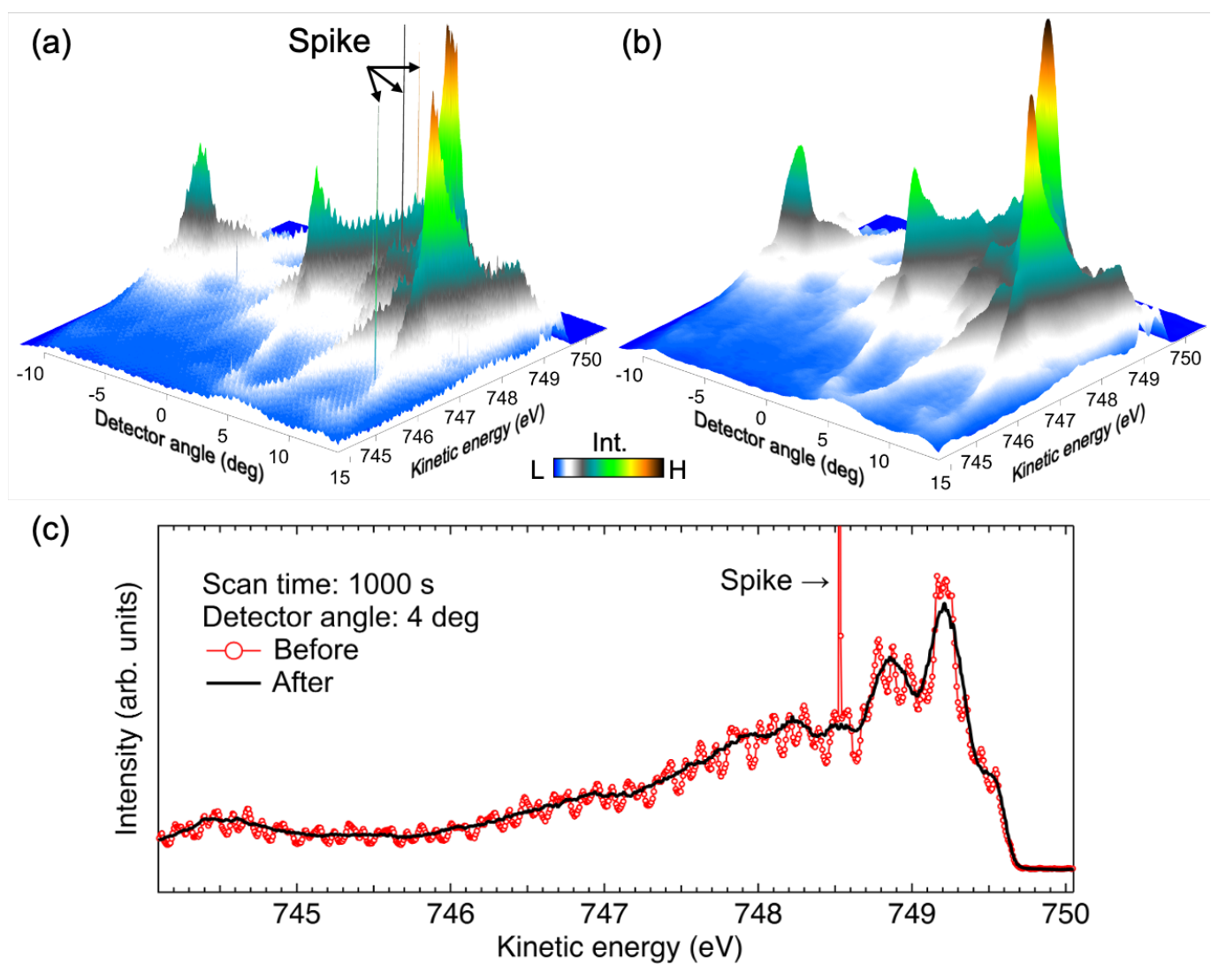}
\caption{\label{fig:epsart} The upper two figures present 3D surface plots of CeRu$_2$Si$_2$ (a) before and (b) after denoising. The data were acquired at $h\nu = 750$~eV and 77~K in Fixed mode with an accumulation time of 1000 s, which ensures a sufficient S/N ratio. (c) Comparison of EDC spectra at a detector angle of 4$^\circ$, before and after denoising, highlighting the removal of instrument artifacts.}
\label{Fig.2}
\end{figure}
\begin{figure*}
\includegraphics[width=16cm]{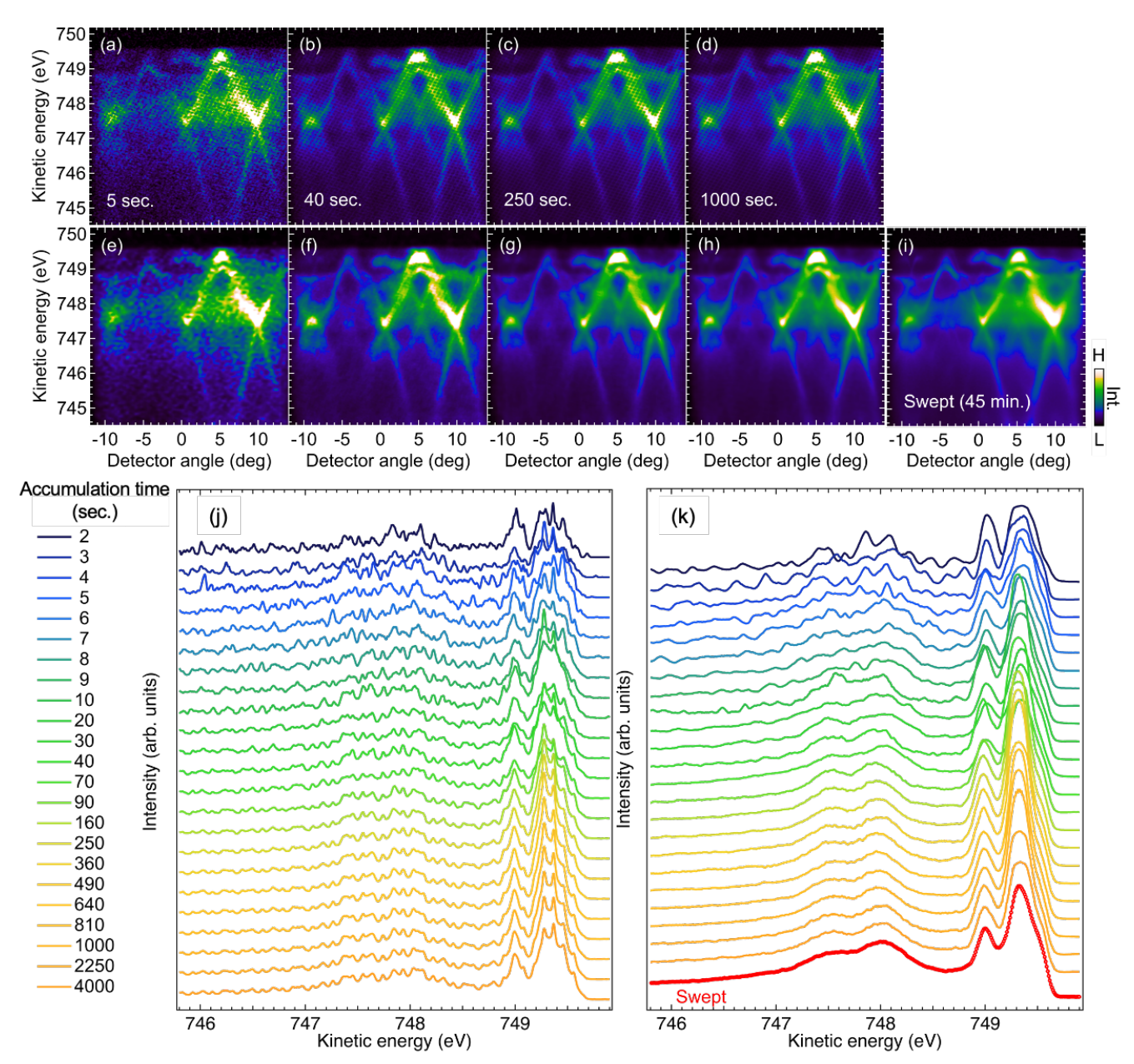}
\caption{Relationship between accumulation time and raw kinetic energy-detector angle ($E_k$-$\theta$) images for CeRu$_2$Si$_2$ acquired in Fixed mode. (a-d) $E_k$-$\theta$ images with different accumulation time. (e-h) The same $E_k$-$\theta$ images as panels (a-d), but after denoising. (i) Raw $E_k$-$\theta$ image measured in Swept mode for comparison, with an accumulation time of 45 minutes. (j-k) Comparison of the accumulation time dependent EDC spectra at detection angle of 5.5$^\circ$ (j) before and (k) after denoising. The EDC spectrum acquired in Swept mode is also shown for comparison.}
\label{Fig.3}
\end{figure*}
\section{Overview of the denoise system and its implementation in the $\mu$SX-ARPES system at SPring-8 BL25SU}
The $\mu$SX-ARPES system operates at the circularly polarized soft X-ray beamline BL25SU of SPring-8~\cite{senba2016upgrade,muro2021soft}.
The available $h\nu$ range is from 120~eV to 2000~eV.
The resolving powers ($h\nu/\Delta h\nu$) of the monochromator typically used range from 10000 for standard ARPES to 20000 for higher energy resolution ARPES.
Using the Wolter mirror, a micro beam spot size is typically achieved on the sample~\cite{senba2020stable,muro2021soft}.
The small footprint of this micro-focused beam allows for measurements in a grazing-angle geometry, with the glancing angle between the beam axis and the sample surface set to approximately 5$^\circ$.
This geometry enhances the experimental efficiency by more than an order of magnitude compared to the conventional 45$^\circ$ configuration~\cite{muro2021soft}.
ARPES data are recorded by the DA30 hemispheric analyzer (Scienta Omicron Inc.).
This analyzer can perform two-dimensional angular ($\theta_x$-$\theta_y$) mapping while maintaining a fixed geometry between the X-ray source, sample, and DA30, by using a deflector scan perpendicular to the detector slit.

Since BL25SU is a public beamline, in-situ processing must be highly practical to facilitate user access to DPDM.
Therefore, we have developed an environment that enables (i) fast processing, (ii) ease of operation, and (iii) a high degree of flexibility.
Figure~\ref{Fig.1}(c) provides an overview of the ultra-high efficiency $\mu$SX-ARPES system environment.
In addition to the PC for measurement and analysis/optics, a dedicated PC for DPDM is utilized, and experimental data are shared through a local network connection.
DPDM-PC is operated remotely from the analysis/optics PC.
The DPDM-PC is equipped with memory (DDR5 4800MHz RDIMM 16GB $\times$ 8), an SSD (ADD-SSD-S3840G), and a GPU (ADD-NV-RTXA6000) to maximize noise processing and image storage capacity.

DPDM~\cite{deep_prior} treats the denoising task as an optimization problem in which the network weights $\theta$ are updated to minimize the loss function $L(\theta)$, defined as MSE between the output $f_{\theta}(z)$ and the observed image $x_{obs}$:$$L(\theta) = \frac{1}{n} \sum_{i=1}^{n} \| f_{\theta}(z)_i - (x_{obs})_i \|^2$$ where $n$ is the total number of pixels.
Prior to this optimization, a pre-processing step is implemented to address intense spike structures, which appears as outliers in the photoelectron intensity distribution.
By clipping these outliers and narrowing the dynamic range of the intensities to be estimated, the convergence of the optimization is significantly improved.
While the intensities of these clipped pixels are initially treated as missing values, DPDM effectively interpolates them by leveraging the structural priors of the CNN.
Unlike a conventional median filter that simply replaces a pixel with the median of its neighbors and can inadvertently degrade valid neighboring signals, DPDM reconstructs the signal based on the physical continuity and global features of the ARPES spectra, ensuring a more robust and high-fidelity restoration of the intrinsic electronic structure.
This method exploits the spectral bias of CNNs~\cite{rahaman2019spectral}, where the network inherently learns low-frequency band structures faster than high-frequency instrumental artifacts.
Consequently, the intrinsic ARPES signal is effectively isolated during the early stages of optimization, particularly within the loss function’s plateau region, before the network begins to overfit the instrumental artifacts.

DPDM is operated using Python code implemented in a Jupyter Notebook environment, as shown in Fig.~\ref{Fig.1}(d).
To enhance user-friendliness, the system requires only the input of two-dimensional numerical data and the designated denoising range.
In our implementation, the total number of iterations and the save span (the interval at which intermediate images are saved) can be manually adjusted to accommodate diverse ARPES datasets with different dispersion profiles and S/N ratios.
Typically, the system saves an image of the ongoing process every 10–50 iterations.
The movie in the Supplementary Information demonstrates the denoising process, which is completed in approximately 30 s.
After execution, the user selects the optimal ARPES image by monitoring the loss function and the corresponding ARPES images displayed in the Jupyter Notebook.
Figure~\ref{Fig.1}(e) presents an ARPES image at a specific iteration alongside the loss function~\cite{deep_prior}.
Denoising is achieved by extracting the intrinsic band structure from the input data, starting from uniformly distributed random noise.
Instrumental artifacts are effectively removed when the loss function reaches a plateau (e.g., around 300 iterations).
Beyond this point (e.g., after 400 iterations), the loss function may decrease further, reflecting overfitting where instrumental artifacts gradually re-emerge.
Therefore, the ARPES image captured immediately after the loss function converges to a constant value is selected as the denoised data.
As demonstrated in Figs.~\ref{Fig.2}(c) and~\ref{Fig.3}, hereafter mentioned below, this protocol strictly preserves the key spectral intensities and structural trends without introducing artificial distortions.

\section{Evaluation of efficiency for $\mu$SX-ARPES with denoise system}
\subsection{Reducing Accumulation Time}
To evaluate the effectiveness of DPDM in removing instrumental artifacts, we first performed $\mu$SX-ARPES measurements on CeRu$_2$Si$_2$ as a demonstration.
CeRu$_2$Si$_2$ is a well-known 4$f$ heavy fermion system.
The three-dimensional band structure and its spectral analysis using energy distribution curves (EDCs) have been reported in previous SX-ARPES studies~\cite{yano2008electronic,okane20094,okane2011resonant}.
All subsequent ARPES data were recorded in Fixed mode at $h\nu = 750$~eV with $h\nu/\Delta h\nu = 10000$ and a temperature of 77~K.
Clean surface was obtained by cleaving under 2.0$\times$10$^{-8}$ Pa at 77~K.
Figures~\ref{Fig.2}(a) and (b) present the 3D surface plots before and after denoising.
The accumulation time was set to 1000 s to ensure a sufficient S/N ratio.
Despite the presence of significant spike structure in the raw data, alongside the band dispersion, this artifact is effectively removed via the pre-processing and interpolation mentioned above.
The periodic grid structure is smoothly eliminated from the EDC spectra (Fig.~\ref{Fig.2}(c)).
Therefore, we conclude that DPDM effectively removes instrumental artifacts, enabling spectral analysis.

Next, we investigated the potential of DPDM to reduce the accumulation time.
Figures~\ref{Fig.3}(a-d) show the accumulation time dependence of raw kinetic energy-detector angle ($E_k$-$\theta$) images acquired in Fixed mode.
While the data acquired with an accumulation time of 5 s exhibit a low S/N ratio due to statistical errors, those acquired with accumulation times longer than 40 s appear as smooth images.
However, after denoising (Figs.~\ref{Fig.3}(e-h)), the band dispersion can be discerned even in the data acquired with a 5-second accumulation time.
Fourier analysis indicates that the high-frequency component originating from the grid structure was effectively removed from both the 10-second and 40-second data~\cite{deep_prior}.
In comparison to the $E_k$-$\theta$ image acquired in Swept mode~\cite{Supplement} (Fig.~\ref{Fig.3}(i)) with a 45-minute accumulation time, we conclude that sufficient S/N ratio data can be obtained with an accumulation time of at least 40 s.
Using Fig.~\ref{Fig.3}(j) and (k), we conducted a more detailed examination of the minimum accumulation time required to achieve an S/N ratio suitable for EDC spectral analysis.
The EDC spectrum obtained in Swept mode reveals that the broad structure observed around $E_k$ = 747.75~eV comprises two peaks.
Using this low-intensity two-peak structure as a benchmark, we determined that the threshold accumulation time required for reliable observation of these features in the denoised Fixed mode data is 40 s.
Therefore, we conclude that statistically reliable SX-ARPES data, equivalent in quality to the Swept-mode data that requires a 2700 s accumulation, can be acquired in only about 70 s (combining 40 s of accumulation and 30 s of DPDM processing).
This represents a 40-fold improvement in measurement efficiency, demonstrating the significant impact of DPDM on accelerating high-resolution SX-ARPES experiments.

\subsection{Applying To The Material With Indistinct Band Structure}
Even when ARPES measurements are performed on an atomically flat and clean surface, the resulting image is not always clear.
For example, the scattering of photoelectrons caused by crystal defects, such as impurity atoms, lattice imperfections, and dislocations, as well as electron-electron and electron-phonon interactions, can obscure the band structure.
These interactions, along with the finite lifetimes of both the photoelectrons and the photo-excited holes, contribute to the intrinsic spectral broadening~\cite{sobota2021angle}.
Furthermore, the uncertainty in the perpendicular momentum of photoelectrons, known as  $k_z$-broadening, which is inversely proportional to the detection depth, contributes to the formation of intrinsic spectral broadening~\cite{sobota2021angle,strocov2003intrinsic,strocov2023high}.
Particularly in the soft X-ray region, the Debye-Waller factor plays a critical role; as temperature increases, the intensity of coherent spectral features decreases, while the incoherent background from thermally diffused scattering increases~\cite{braun2013exploring}.
To address these challenges, we demonstrate the effectiveness of DPDM in extracting band structures obscured by these complex crystallographic, many-body, and thermal effects.
\begin{figure}
\includegraphics[width=8.5cm]{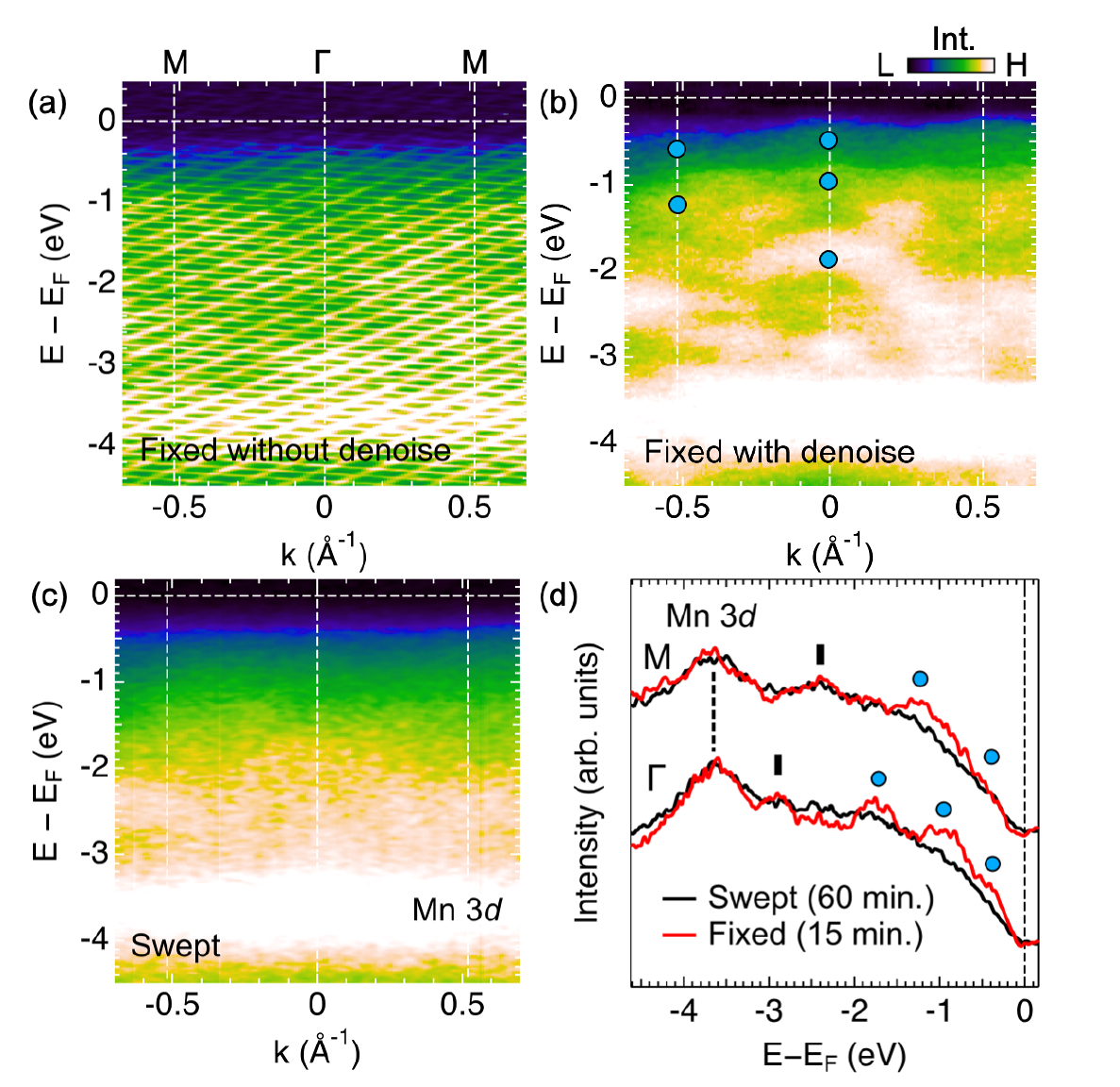}
\caption{The binding energy - momentum ($E$-$k$) images of Mn$_3$Si$_2$Te$_6$ acquired in Fixed mode (a) without and (b) with denoising. (c) The same $E$-$k$ image, but obtained in Swept mode for comparison. The data were acquired at $h\nu = 500$~eV and a temperature below 30~K along the $\Gamma$-$M$ line. (d) EDC spectra at the $\Gamma$ and $M$ points. The filled circles indicate the peak clearly observed in Fixed mode, but not in Swept mode.}
\label{Fig.4}
\end{figure}

We employed a ferrimagnetic semiconductor Mn$_3$Si$_2$Te$_6$ as a test material~\cite{may2017magnetic,ni2021colossal,seo2021colossal}.
Previous VUV-ARPES measurements have shown that the band dispersion is obscured by spectral broadening due to $k_z$ projection, as well as a significant background~\cite{bigi2023covalency}.
SX-ARPES data were acquired at $h\nu = 500$~eV with $h\nu/\Delta h\nu = 10000$ and below 30~K along $\Gamma$-$M$ line within the Brillouin Zone.
Low temperature mitigate the reduction of coherent spectral weight caused by the Debye-Waller factor.
Clean surface was obtained by cleaving under 2.0$\times$10$^{-8}$ Pa below 30~K.
The accumulation time was set to 1000 s for Fixed mode and 3600 s for Swept mode.
Figures~\ref{Fig.4}(a) and (b) present the binding energy - momentum ($E$-$k$) image obtained in Fixed mode without and with denoising, respectively.
Instrumental artifacts are effectively removed, and multiple band structures are extracted, resulting in sharper image contrast.
As shown in Fig.~\ref{Fig.4}(c) for comparison in Swept mode, in addition to a Mn 3$d$ flat band at $E-E_F = -3.6$~eV and a convex band dispersion around $E-E_F = -2.5$~eV, the band dispersion around $E-E_F = -1.0$~eV is more clearly resolved in Fixed mode with denoising.
This image sharpening corresponds to the enhanced peak definition in the EDC spectrum (Fig.~\ref{Fig.4}(d)).

Compared to VUV-excited photoelectrons, SX-excited photoelectrons have larger escape depth and kinetic energy, which reduces the intrinsic $k_z$-broadening and the background arising from electronic scattering.
Furthermore, the increased bulk sensitivity in the SX region, combined with the suppressed thermal diffuse scattering at low temperature, allows for a clearer observation of the intrinsic electronic structure.
However, the relatively blurred images and broadened EDC spectra observed in Swept mode indicate that the potential physical advantages are obscured by several factors~\cite{Supplement}.
In addition to instrumental instabilities and time-dependent surface degradation during the extended acquisition, the instrumental background from the voltage sweep and accumulated artifacts further hinder peak formation.
Therefore, we conclude that DPDM, in conjunction with the advantages of the Fixed mode, effectively eliminates instrumental artifacts to reveal the intrinsic EDC spectra.
\begin{table}
  \caption{List of the detectable $E_k$ range selected by operating mode and $E_p$ in the case of DA30 analyzer at BL25SU in SPring-8.}
  \vspace{8mm}
  \begin{tabular}{ccc}
    \hline
    \hline
    \\
    \ \ Lens mode \ \ &\ \ DA30-01\ \ &\ \ DA14-08\ \ \\
    \ \ Detector angle\ \ &\ \ $|\theta_x|\le 15{^\circ}$, $|\theta_y|\le 10{^\circ}$\ \ &\ \ $|\theta_x|\le 7{^\circ}$, $|\theta_y|\le 7{^\circ}$\ \ \\
    \\
    \hline
    \ \ $E_p$ (eV)\ \ &\multicolumn{2}{c}{$E_k$ (eV)}\\
    \hline
    \\
    100&12-800&22-1071\\
    \\
    75&9-600&18-1167\\
    \\
    50&6-400&13-1131\\
    \\
    \hline
    \hline
  \end{tabular}
\label{TableI}
\end{table}
\section{Discussions: future applications and developments}
The aforementioned demonstration confirms that DPDM successfully reduces the accumulation time required to obtain high S/N spectra, without compromising the performance of the $\mu$SX-ARPES system.
This presents a valuable opportunity to further develop the capabilities of $\mu$SX-ARPES and establish new measurement techniques

\begin{figure}
\includegraphics[width=8.5 cm]{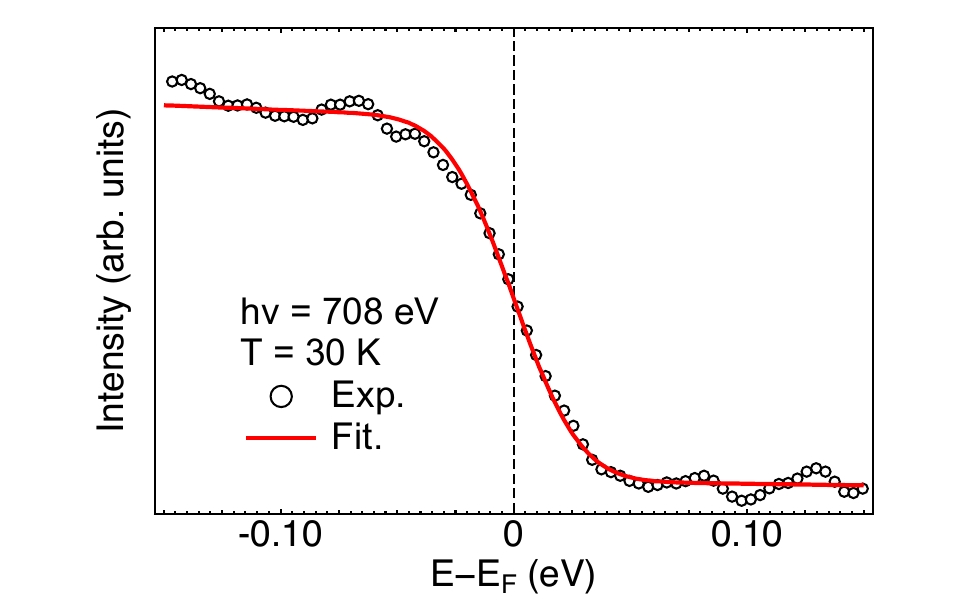}
\caption{Angle-integrated ARPES spectrum near the Fermi level of gold in Swept mode. The red solid line represents the fit of the Fermi-Dirac distribution, with a total energy resolution of 51.6~meV, to the data.}
\label{Fig.5}
\end{figure}

The first opportunity lies in enhancing the overall experimental energy resolution $\Delta E_T$.
Energy resolution has followed a continuous trend of exponential improvement reported in representative SX-ARPES studies~\cite{sobota2021angle}.
$\Delta E_T$ value is basically determined by the combined resolutions of the photon source ($\Delta h\nu$) and the photoelectron analyzer ($\Delta E_A$): $$(\Delta E_T)^2=(\Delta h\nu)^2+(\Delta E_A)^2$$
Here $\Delta E_A\approx\dfrac{w}{2R}E_p$, and $w$, $R$, $E_p$ represent the entrance slit width, hemispherical radius, and pass energy, respectively~\cite{hadjarab1985image,tusche2019imaging}.
For the DA30 analyzer at BL25SU of SPring-8, the standard set values are DA30-01 lens operating mode with $w = 0.2$~mm, $R = 200$~mm, and $E_p = 100$~eV, corresponding to $\Delta E_A = 50$~meV.
However, with decreasing $E_p$ to 50~eV, the detectable kinetic energies of photoelectron become narrower to 400~eV.
In contrast, as shown in Table~\ref{TableI} about the list of operating mode for DA30-01 mode ($|\theta_x| \le 15{^\circ}$, $|\theta_y| \le 10{^\circ}$), DA14-08 mode ($|\theta_x| \le 7{^\circ}$, $|\theta_y| \le 7{^\circ}$) with a narrower detector angle can measure $\theta_x$-$\theta_y$ mapping for kinetic energies up to 1131~eV even at $E_p = 50$~eV.
Actually we measured ARPES on polycrystalline gold in DA14-08 lens mode by Swept mode.
The condition was set to $h\nu = 708$~eV with $h\nu/\Delta h\nu = 20000$, $w = 0.3$~mm, $E_p = 50$~eV, corresponding to $\Delta E_T = 51.6$~meV as theoretical value.
As shown in Fig.~\ref{Fig.5}, we confirmed that $\Delta E_T$ at 30~K is achieved 51.6~meV fitted by Fermi-Dirac distribution to the angle-integrated experimental data with an accumulation time of 2 hours.
This energy resolution is improving following an exponential trend of previous SX-ARPES studies~\cite{sobota2021angle}.
Consequently, the ability of DPDM to achieve high S/N ratios in Fixed mode significantly enhances the feasibility of high-resolution measurements in the soft X-ray region.

At fourth-generation synchrotron radiation facilities such as NanoTerasu, the significantly reduced emittance leads to a smaller source divergence~\cite{tavares2018commissioning,fornek2019advanced,raimondi2021commissioning,klysubun2022sps,watanabe2023spring,obara2025commissioning,horiba2022design}.
In the case of the NanoTerasu ARPES beamline, this low-divergence source is highly compatible with the collimated plane grating monochromator (cPGM) system~\cite{follath1997new, horiba2022design}.
The cPGM allows for a flexible choice of the magnification factor, $c_{ff} = \cos\beta/\cos\alpha$ (where $\alpha$ and $\beta$ are the incident and diffraction angles at the grating, respectively).
By effectively utilizing the well-collimated beam from the low-divergence source and selecting a high $c_{ff}$ value, the beamline can maximize energy dispersion while minimizing optical aberrations, facilitating a high energy resolution ($h\nu/\Delta h\nu \ge 30,000$)~\cite{horiba2022design}.
While narrowing the entrance and exit slits of the monochromator to achieve such extreme resolution inevitably limits the total photon flux on the sample, the high brilliance of the source ensures that a workable count rate is maintained for ARPES measurements.
In this regime, DPDM becomes an essential tool to restore clear spectral features from the resulting low-statistics data.
For example, when $h\nu/\Delta h\nu = 50000$, $w = 0.2$~mm, and $E_p = 50$~eV in DA30 analyzer, $\Delta E_T$ is expected to 29.7~meV at $h\nu = 800$~eV, which can become comparable to that of VUV-ARPES.
This would enable the investigation of three-dimensional momentum-resolved fine electronic structures with bulk sensitivity such as High-T$_{\rm c}$ superconducting cuprates~\cite{damascelli2003angle,matt2018direct,horio2018three,yu2020relevance}.

The second opportunity is the development of coherent synchrotron radiation SX-ARPES for the observation of three-dimensional nonequilibrium electron structures.
SPring-8-II plans to achieve an emittance reduction of more than one order of magnitude compared to SPring-8, in the short wavelength region above hard X-ray~\cite{watanabe2023spring}.
This suggests the possibility of utilizing fully coherent synchrotron radiation in the soft X-ray region.
The use of temporal coherence offers significant value to the nonequilibrium studies, as reported by the advancements in time-resolved soft X-ray measurements ~\cite{stamm2007femtosecond,yamamoto2013time,higley2016femtosecond,takubo2017capturing,kraus2018ultrafast,yokoyama2019photoinduced,ismail2020direct,zhang2022photo,yamagami20244}.
Time-resolved ARPES has been pioneered using ultraviolet and extreme ultraviolet laser-based sources~\cite{zhou2018new,puppin2019time,lv2019angle,madeo2020directly,suzuki2021hhg,na2023advancing,boschini2024time}, and soft X-ray sources are crucial for extending these studies to bulk electronic structures in a wider momentum space.
The developed DPDM, in conjunction with next-generation synchrotron soft X-ray light sources, is expected to dramatically advance ARPES technology.

\section{Summary}
We have established an ultra-efficient SX-ARPES measurement environment by integrating deep-prior based denoise system into the $\mu$SX-ARPES system at BL25SU in SPring-8.
The challenge posed by the limited photoionization cross-section was overcome by reducing both the denoising processing time and the accumulation time in Fixed mode on the photoelectron analyzer.
This also enables the visualization of band dispersion obscured by crystallographic electron scattering and electron correlation.
We anticipate that next-generation synchrotron radiation sources and advancements in the electrical control of photoelectron analyzers will lead to the development of enhanced ARPES technology, facilitating further improvements in energy resolution and enabling the exploration of three-dimensional nonequilibrium electronic structures.
Furthermore, the utility of DPDM extends beyond SX-ARPES; it can be readily applied to VUV-ARPES images afflicted by instrumental artifacts.
The technique is broadly anticipated to be highly effective for a diverse range of experimental image data, spanning applications from periodic background removal in data to the analysis of reflective images in X-ray diffraction, photoelectron diffraction patterns in photoelectron holography, and speckle patterns in X-ray photon correlation spectroscopy.

\section*{Supplementary Material}
See the supplementary material for the movie about the denoising program and the Python code and manual of DPDM system, along with demonstration $\mu$SX-ARPES data for CeRu$_2$Si$_2$ accumulated over 1000 s.

\begin{acknowledgments}
We thank for T. Ohkochi and Y. Amakai for providing the single crystal sample.
The soft X-ray ARPES experiments were performed at the BL25SU of SPring-8 with the approval of the Japan Synchrotron Radiation Research Institute (JASRI) (Proposal No.2023A1354, 2023B2426, 2024A2397, 2024B2440, 2025A2384).
This work was supported by Grant-in-Aid for Early-Career Scientists (Grant No.25K17944) and JST PRESTO (Grant No. JPMJPR25JA).
\end{acknowledgments}

\bibliography{aipsamp}

\end{document}